\documentclass[graybox]{svmult}

\usepackage{mathptmx}       % selects Times Roman as basic font
\usepackage{helvet}         % selects Helvetica as sans-serif font
\usepackage{courier}        % selects Courier as typewriter font
\usepackage{type1cm}        % activate if the above 3 fonts are
\usepackage{makeidx}         % allows index generation
\usepackage{graphicx}        % standard LaTeX graphics tool
\usepackage{multicol}        % used for the two-column index
\usepackage[bottom]{footmisc}% places footnotes at page bottom

\def\rmit#1{{\it #1}}              %% italics (RR mode, Kluwer)

\def\etal{\rmit{et al.}}

\def\eg{\rmit{e.g.}}

\makeindex             % used for the subject index
\begin{document}

\title*{Magnetic fingerprints of solar and stellar oscillations}
\titlerunning{Magnetic fingerprints of solar and stellar oscillations}
\author{Elena Khomenko\inst{1}}
\institute{Instituto de Astrof\'{\i}sica de Canarias, 38205, C/
V\'{\i}a L{\'a}ctea, s/n, Tenerife, Spain; \\
 Main Astronomical Observatory, NAS, 03680, Kyiv, Ukraine; \texttt{khomenko@iac.es}}
\maketitle
\index{Author1}

 \abstract{Waves connect all the layers of the Sun,
from its interior to the upper atmosphere. It is becoming clear
now the important role of magnetic field on the wave propagation.
Magnetic field modifies propagation speed of waves, thus affecting
the conclusions of helioseismological studies. It can change the
direction of the wave propagation, help channeling them straight
up to the corona, extending the dynamic and magnetic couplings
between all the layers. Modern instruments provide measurements of
complex patterns of oscillations in solar active regions and of
tiny effects such as temporal oscillations of the magnetic field.
The physics of oscillations in a variety of magnetic structures of
the Sun is similar to that of pulsations of stars that posses
strong magnetic fields, such as roAp stars. All these arguments
point toward a need of systematic self-consistent modeling of
waves in magnetic structures that is able to take into account the
complexity of the magnetic field configurations. In this paper, we
describe simulations of this kind, summarize our recent findings
and bring together results from the theory and observations.}

\section{Introduction}

%Waves generated y turbulence inside the stars: consequences and
%wave properties according to theoretical and numerical results.

%Waves resonate in the cavity and heat the chromosphere: briefly
%problems of the chromospheric heating theory. Hope:  magnetic
%field. Problems of helioseismology: problems caused by the
%magnetic field.

%Magnetic field affects the properties of waves: in which way. New
%wave types, properties and mode conversion.

Any turbulent medium, as the interior of the Sun or stars,
generates sound. The basics of the theory excitation of sound
waves by the turbulent flow were developed by Lighthill in 1952
\cite{Lighthill1952}. Since then, a vast amount of theoretical and
numerical efforts has been dedicated to specify the properties of
the spectrum of sound waves generated in a stratified stellar
convection zone, by improving the description of the turbulent
energy spectrum of the convective elements \eg\ \cite{Bi+Li1998,
Goldreich+Kumar1990, Goldreich+Murray+Kumar1994,
Musielak+etal1994, Nordlund+Stein2001, Stein1967,
Stein+Nordlund2001}. Without going into details of these works,
the present knowledge can be summarized in the following way. The
efficiency of the energy conversion from convective to acoustic is
proportional to a high power of the Mach number of the convective
motions ($M^{15/2}$ \cite{Goldreich+Kumar1990}).  Most of the
energy going into the $p$-modes, $f$-modes and propagating
acoustic waves is emitted by convective eddies of size $h \sim
M^{3/2}H$ (where $H$ is the value of the pressure scale height),
at frequencies close to the acoustic cut-off frequency  $ \omega
\sim \omega_c$ and wavelengths similar to $H$. This defines the
frequency dependence of the oscillation spectrum observed in the
Sun and stars \cite{Goldreich+Kumar1990}.
Since the Mach number is largest at the top part of the convection
zone, the peak of the acoustic energy generation is located
immediately below the photosphere \cite{Nordlund+Stein2001,
Stein+Nordlund2001}.
Recent numerical simulations of magneto-convection have shown that
the magnetic field modifies the spectrum of waves
\cite{Jacoutot+etal2008}. Apart from the power suppression in
regions with enhanced magnetic field, these simulations suggest an
increase of high-frequency power (above 5 mHz) for intermediate
magnetic field strengths (of the order of 300--600 G) caused by
changes of the spatial-temporal spectrum of turbulent convection
in a magnetic field.

%Lee (1993, ApJ, 404, 372) for generation of waves in a sunspot.

Waves generated in the convection zone resonate inside a cavity
formed by the stellar interior and the photosphere and are used by
helio- and astroseismology to derive its properties
\cite{Deubner1975, Ulrich1970}. The information contained in the
frequencies of the trapped wave modes is used by the classical
helioseismology. A relatively newer branch called local
helioseismology uses the information contained in the velocity
amplitudes of the propagating waves
%(supposed they are just produced by the source and did not travel
%enough inside the Sun to become trapped)
measured in a region of interest on the solar surface (Duvall
\etal\ \cite{Duvall+etal1993}). By inversion of these
measurements, variations of the wave speed and velocities of mass
flows can be recovered below the visible solar surface.
The inversion results have been obtained for quiet Sun regions as
well as for magnetic active regions including sunspots. It is
known that sunspots possess strong magnetic fields with a
complicated structure in the visible layers of the Sun where the
Doppler measurements used by local helioseismology are taken
\cite{Solanki2003}. Consequently, such magnetic field can cause
important effects on helioseismic waves, beyond the perturbation
theories employed for helioseismic data analysis
\cite{Birch+Kosovichev2000, Gizon+Birch2002,
Kosovichev+Duvall1997}.
Recent numerical and analytical results demonstrate that the
observed time-distance helioseismology signals in sunspot regions
correspond to fast MHD waves \cite{Khomenko+etal2008b,
Moradi+etal2008}.

An estimation of the acoustic energy flux generated in the solar
convection zone by the turbulent motions suggests that the it can
be as large as, \eg\ $F_A = 5 \times 10^7$ ergs/cm$^2$/s
\cite{Musielak+etal1994}. This is more than sufficient to maintain
a hot chromosphere. It made the theory of acoustic heating of the
upper atmosphere very attractive.
However, soon after being proposed, the theory of acoustic heating
encountered several major difficulties. It was found that both,
low- and high-frequency waves are radiatively damped in the
photosphere, reaching the upper layers with significantly less
power \cite{Ulmschneider1971}.  An additional difficulty comes
from the fact that the measured high-frequency acoustic fluxes in
the photosphere and chromosphere are uncertain and non-conclusive
\cite{Kalkofen2007, Wunnenberg+etal2002}. Low-frequency acoustic
waves are reflected in the photosphere due to the effects of the
cut-off frequency ( $\sim 4$ mHz) before reaching chromospheric
heights. Due to their long wavelengths they have too large shock
formation distances.
Despite that, the five-minute waves with enough energy were
detected in the chromosphere and corona of the Sun  mainly above
solar facular and network areas \cite{Centeno+etal2006b,
Krijer+etal2001, DeMoortel+etal2002, DePontieu+etal2003,
Veccio+etal2007}. Several explanations of these waves involving
the magnetic field have been recently proposed
\cite{Khomenko+etal2008, DePontieu+etal2004}.
The wave energy can reach the upper layers not necessarily in the
form of acoustic waves, but also in the form of other wave types,
like magneto-acoustic waves, Alfv\'en waves or a family of waves
propagating in thin magnetic flux tubes (see a recent example in
\eg\ \cite{Hindman+Jain2008}). All of them are related to the
magnetic field structure. Osterbrock in 1961 \cite{Osterbrock1961}
was the first to incorporate magnetic field into the theory of
wave heating and to point out its importance on the propagation
and refraction characteristics of the fast and slow MHD waves.

The above examples are only a few where the influence of the
magnetic field on the wave properties is demonstrated to be
important. Magnetic field not only changes the acoustic excitation
rate and produces new wave modes. It also modifies the wave
propagation paths and the direction of the energy propagation, it
produces wave refraction and changes the reflection
characteristics at the near-surface layers. Magnetic field defines
the wave propagation speeds and changes the acoustic cut-off
frequency. It can also change the wave dissipation rates and
provides an additional energy source. Magnetic field connects all
the atmospheric layers facilitating the channeling of waves from
the lower to the upper atmosphere. This makes the magnetic field
an important ingredient the theories of the wave propagation in
the atmospheres of the Sun and stars.

Apart from the problems set by the local helioseismology in
magnetic regions and the wave heating theory, of pure physical
interest is the interpretation of oscillations observed in
different magnetic structures in terms of MHD waves. For example,
the wave dynamics seen in high-resolution DOT movies of a sunspot
region \cite{Rouppe+etal2003} demonstrate that phenomena such as
chromospheric umbral flashes and running penumbral waves are
closely related. What type of waves are responsible for them?
Solar small-scale and large-scale magnetic structures have
distinct magnetic and thermal properties and support different
wave types. The observed frequency spectrum of waves in these
structures is not the same (see the introduction in
\cite{Khomenko+Collados2007}). Numerical simulations of waves in
non-trivial magnetic structures (\eg\ \cite{Bogdan+etal2003,
Hasan+Ballegooijen2008, Hasan+etal2005, Khomenko+Collados2006,
Khomenko+Collados2007, Rosenthal+etal2002}) have shown the complex
pattern formed by waves of various types that can propagate
simultaneously in various directions.
Different wave modes can be detected in observations depending on
the magnetic field configuration and the height where acoustic
speed, $c_S$, and Alfv\'en speed, $v_A$, are equal relative to the
height of formation of the spectral lines used in observations.

During the last years we applied efforts to develop a numerical
code aimed at calculating the non-linear wave propagation inside
magnetic fields in 2 and 3 dimensions. Using this code we focused
our analysis on several problems described above, namely:
magneto-acoustic wave propagation and refraction in sunspots and
flux tubes; channeling the five-minute photospheric oscillations
into the solar outer atmosphere through small-scale magnetic flux
tubes; influence of the magnetic field on local helioseismology
measurements in active regions. The results of these studies are
published in \cite{Khomenko+etal2008, Khomenko+Collados2006,
Khomenko+Collados2007, Khomenko+etal2008b}. In the rest of the
paper we briefly summarize the results and conclusions of these
works. In addition, the last section gives our recent contribution
to the problem of the interpretation of observations of waves in
magnetic roAp stars.

%%%%%%%%%%%%%%%%%%%%%%%%%%%%%%%%%%%%%%%%%%%%%%%%%%%%%%%%%%%%%%%%%%%%%%%%
\begin{figure*}
\includegraphics[width=12cm]{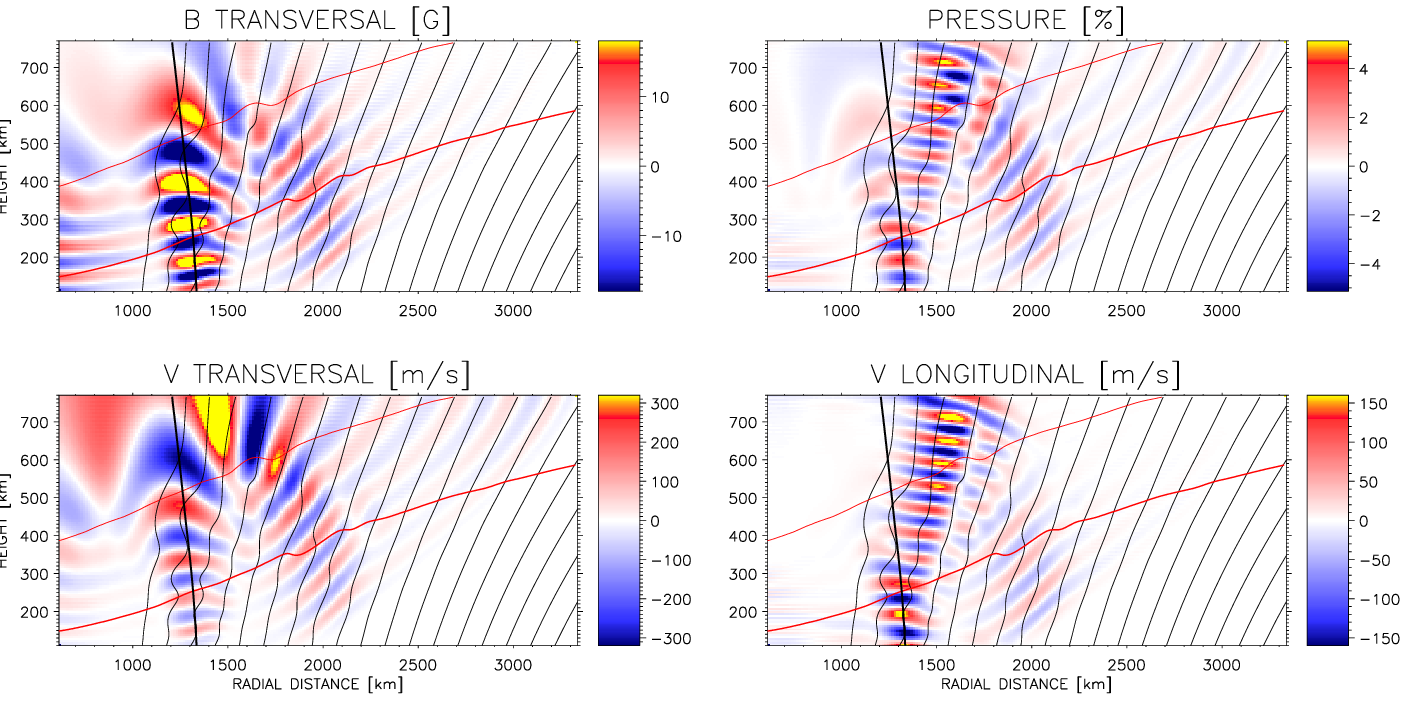}
\caption{Variations of the velocity, magnetic field and pressure
at an elapsed time $t=100$ sec after the beginning of the
simulations. In each panel, the horizontal axis is the radial
distance $X$ from the sunspot axis and the vertical axis is height
from the photospheric level. The black inclined lines are magnetic
field lines. The two more horizontal lines are contours of
constant $c_S^2/v_A^2$, the thick line corresponding to $v_A=c_S$
and the thin line to $c_S^2/v_A^2=0.1$. Top left: transversal
variations of the magnetic field. Top right: relative pressure
variations. Bottom panels: transversal and longitudinal variations
of the velocity. \label{fig:long}}
\end{figure*}
%%%%%%%%%%%%%%%%%%%%%%%%%%%%%%%%%%%%%%%%%%%%%%%%%%%%%%%%%%%%%%%%%%%%%%%%

\section{Waves in sunspots}

Sunspots show a very complex dynamics as revealed by
high-resolution observations. The umbral flashes observed in the
chromosphere are thought to be acoustic shock waves
\cite{Rouppe+etal2003} propagating along the nearly vertical
magnetic field lines. The visible perturbation then expands
quasi-circularly out to the penumbra producing the running
penumbral waves \cite{Tziotziou+etal2006}. This visible pattern
can be interpreted as slow (acoustic) wave propagating along the
inclined magnetic field lines in the penumbra
\cite{Bloomfield+etal2007}. The power distribution at different
frequencies in active regions is rather complex as well.
The recent  observations of HINODE \cite{Nagashima+etal2007}
confirmed the conclusions done with the ground-based observations
(see \eg\ Tziotziou \cite{Tziotziou+etal2007}) that the dominating
frequency of oscillations within a sunspot depends on the position
in the umbra or penumbra, as well as on the height in the
atmosphere. In the chromosphere the greatest power is observed in
the umbra at 5--6 mHz and a sharp transition between umbra and
penumbra where the dominating frequency is 3 mHz. In the umbral
photosphere the oscillation power is generally suppressed compared
to the quiet photosphere, except for the excess of low-frequency
power at 1-2 mHz at the umbra-penumbra boundary.
An unknown question is what drives the oscillations observed in
sunspots? Are these waves due to a resonant response of the
sunspot flux tube to the external driving by $p$-modes? Are there
sources of oscillations inside the magnetized regions?

%The characteristic speeds of waves change many orders of magnitude
%along the photosphere and chromosphere of sunspots. At some height
%waves propagate through the region where the sound and the
%Alfv\'en speed are equal ($c_S = v_A$). In this region mode
%transformation and coupling of different wave phenomena occurs. In
%addition, the magnetic field and thermodynamic variables show
%important gradients both in horizontal and vertical directions.
%All these ingredients make realistic modeling of sunspot waves a
%rather difficult task only accessible via numerical simulations.

In the simulations described below we aim at identifying the types
of waves modes observed in different layers of a sunspot
atmosphere. We supposed a source of high-frequency (100 mHz)
monochromatic waves located at photospheric level inside the umbra
where the acoustic speed is slightly larger than the Alfv\'en
speed. The waves are assumed to be linear. The initial unperturbed
magnetostatic sunspot model was evaluated following the strategy
described in \cite{Pizzo1986} with a maximum magnetic field
strength of 2.2 kG and characteristic radius of 6 Mm. The
simulation domain is 2-dimensional, extending 0.86 Mm in the
vertical and 3.5 Mm in the horizontal directions.

A snapshot of the simulations is given in Fig.~\ref{fig:long}. The
source generates a set of fast magneto-acoustic waves propagating
upwards. After the perturbation reaches the height where $c_S =
v_A$ the mode transformation takes place. The fast (acoustic) mode
propagating vertically below $c_S = v_A$ level is transmitted as
fast (magnetic) mode in the $c_S < v_A$ region (visible in the
transversal velocity). Its wavelength increases with height due to
the rapid increase of the Alfv\'en speed. The left part of the
wave front of the fast mode (closer to the axis) propagates faster
than its right part (farther from the axis) which produces its
reflection back to the photosphere at some height above the $c_S =
v_A$ level.
The same happens to the other fast mode propagating to the left,
except that this fast wave refracts toward the sunspot axis.
A part of the original fast (acoustic) mode energy is transformed
to the slow (acoustic) mode in the $c_S < v_A$ region. The slow
mode is visible in the longitudinal velocity snapshots. It
propagates with a lower speed, close to the local speed of sound.
This slow mode is channeled along the magnetic field lines higher
up to the chromosphere increasing its amplitude. A similar
behaviour is observed in simulations with an initial condition in
the form of an instantaneous pressure pulse rather than a
monochromatic wave. From these simulations we conclude that in
sunspots only a small fraction of the energy of the photospheric
pulse can be transported to the upper layers, since an important
part of the energy is returned back to the photosphere by the fast
mode. Above a certain height in the low chromosphere, only slow
(acoustic) modes propagating along the magnetic field lines can
exist in the umbra defining the dominating frequency of waves
according to the cut-off frequency of the sunspot atmosphere.

\section{Waves in small-scale flux tubes}

%Though large-period waves are observed energetic enough in
%spicules. How can they be transported there?

%Fawzy et al 2002, 2003, Ulmschneider et al 1971, 2005, Fossum \&
%arlsson (2006) - heating by the short period waves. Carlsson &
%Stein 1997 (simulations of H and K bright points). Kalkoffen et al
%1999 (full-time hot chromosphere or not).

%One figure, channeling of waves vertically upwards. Importance of
%the radiative relaxation time.

%%%%%%%%%%%%%%%%%%%%%%%%%%%%%%%%%%%%%%%%%%%%%%%%%%%%%%%%%%%%%%%%%%%%
\begin{figure}[t]
\centering
\includegraphics[width=12cm]{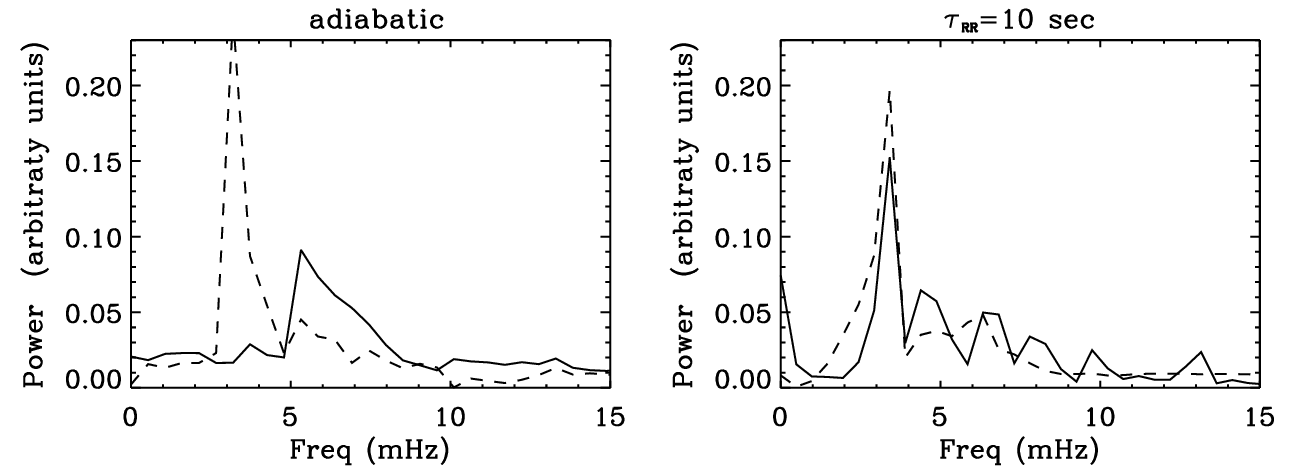}
\caption{Left: photospheric at 400 km (dashed line) and
chromospheric at 1500 km (solid line) power spectra for adiabatic
simulations of waves in flux tubes. Right: same but for simulation
with radiative relaxation time $\tau_{RR}$ equal to 10 s.}
\label{fig:spectras}       % Give a unique label
\end{figure}
%%%%%%%%%%%%%%%%%%%%%%%%%%%%%%%%%%%%%%%%%%%%%%%%%%%%%%%%%%%%%%%%%%%%

Acoustic waves propagating vertically in the quiet solar
atmosphere change their dominating frequency with height from 3
mHz in the photosphere to 5 mHz in the chromosphere. An
explanation for this effect was suggested by Fleck \& Schmitz
\cite{Fleck+Schmitz1991}, who argue that this is a basic
phenomenon due to resonant excitation at the atmospheric cut-off
frequency. The low temperatures of the high photosphere give rise
to a cut-off frequency around 5 mHz. However, numerous
observations suggest that there is no such change for waves
observed in chromospheric and coronal heights above solar facular
and network regions \cite{Centeno+etal2006b, Krijer+etal2001,
DeMoortel+etal2002, DePontieu+etal2003, Veccio+etal2007}. What are
the mechanisms that allow the 3 mHz waves (evanescent in the
photosphere) to propagate up to chromospheric heights? De Pontieu
\etal\ \cite{DePontieu+etal2004} suggest that the inclination of
magnetic flux tubes in facular regions is essential for the
leakage of $p$-modes strong enough to produce the dynamic jets
observed in active region fibrils.
However, this mechanism is hardly to be at work in the
photosphere, where the plasma $\beta$ is larger than 1 and the
acoustic waves do not have a preferred direction of propagation
defined by the magnetic field. In addition, it can not easily
explain the observations of vertically propagating 3 mHz waves in
facular and network regions at chromospheric heights.
Alternatively, a decrease in the effective acoustic cut-off
frequency can be produced by radiative energy losses in thin flux
tubes (Roberts \cite{Roberts1983}).

We explored the latter mechanism and extended the theoretical
analysis of Roberts \cite{Roberts1983} by means of non-adiabatic,
non-linear 2D numerical simulations of magneto-acoustic waves in
small-scale flux tubes with a realistic magnetic field
configuration. The simulations are obtained by introducing a 3 mHz
harmonic vertical perturbation at the axis of a magneto-static
flux tube. Radiative losses were taken into account by means of
Newton's law of cooling with a fixed value of the radiative
relaxation time $\tau_{RR}$. We compare two identical simulation
runs that differ only by the value of $\tau_{RR}$. The first run
is in adiabatic regime ($\tau_{RR} \rightarrow \infty$) and the
second run has $\tau_{RR}=10$ s constant through the whole
atmosphere.
The magnetostatic flux tube model is constructed after the method
of Pneuman \etal\ \cite{Pneuman+etal1986} with a maximum field
strength of 740 G and radius of 100 km in the photosphere. The
simulation box extends 2 Mm in the chromosphere. The details of
these simulations are explained in \cite{Khomenko+etal2008,
Khomenko+Collados2007}.

The vertical photospheric driver generates a fast magneto-acoustic
wave. This wave propagates upwards through the $c_S=v_A$ layer,
preserving its acoustic nature and being transformed into a slow
magneto-acoustic wave higher up. An essential feature of these
simulations is that the wave perturbation remains almost complete
within the same flux tube. Thus, it can deposit effectively the
energy of the driver into the chromosphere.
The power spectra of oscillations resulted from simulations at two
heights in the photosphere and the chromosphere are displayed in
Fig.~\ref{fig:spectras}. In the adiabatic case (left panel) there
is a shift of the dominating  frequency in the oscillation spectra
from 3 mHz in the photosphere to 5 mHz in the chromosphere. In
contrast, there is no such shift in the case of $\tau_{RR}=10$ s
(right panel). The latter power spectra look very similar to those
obtained from spectropolarimetric observations of a facular region
by Centeno \etal\ \cite{Centeno+etal2006b}.
It confirms that radiative losses play an important role in
small-scale magnetic structures, such as those present in facular
regions and are able to decrease the cut-off frequency and modify
the transmission properties of the atmosphere.

%%%%%%%%%%%%%%%%%%%%%%%%%%%%%%%%%%%%%%%%%%%%%%%%%%%%%%%%%%%%%%%%%%%%%%%%%%%%%%%%%%%%%%
\begin{figure*}[t]
\centering
\includegraphics[width=12cm]{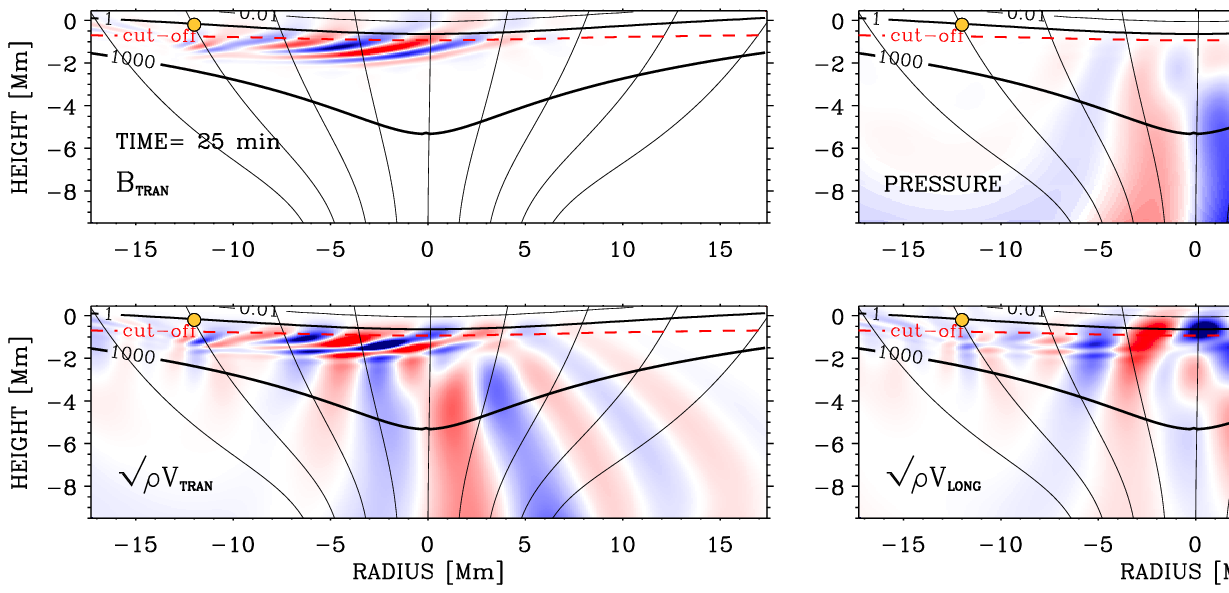}
\caption{Top left: transversal magnetic field variations. Top
right: pressure variations. Bottom panels: transversal and
longitudinal velocity variations. The snapshot is taken 25 min
after the start of the simulations with $B_{\rm phot}$ =1.5 kG.
The dots mark the location of the source. The velocities are
scaled with a factor $\sqrt{\rho_0}$. Black contours of constant
$c_S^2/v_A^2$ are marked with numbers. The dashed contour marks
the position of the cut-off height for 3 mHz waves. Black inclined
lines are magnetic field lines. Negative heights correspond to
sub-photospheric layers. } \label{fig:magnetic}
\end{figure*}
%%%%%%%%%%%%%%%%%%%%%%%%%%%%%%%%%%%%%%%%%%%%%%%%%%%%%%%%%%%%%%%%%%%%%%%%%%%%%%%%%%%%%%%

\section{Local helioseismology in magnetic regions}

Time-distance helioseismology is a branch of local helioseismology
that makes use of wave travel times measured for wave packets
traveling between various points on the solar surface through the
interior. The inversion of these measurements is done under the
assumption that the variations of the travel times are caused by
mass flows and wave speeds below the surface
\cite{Duvall+etal1993, Kosovichev1999, Kosovichev2002,
Kosovichev+etal2000, Zhao+Kosovichev2003}. The interpretation of
results of time-distance seismology encountered major critics when
applied to magnetic active regions of the Sun. Magnetic field in
active regions modifies the wave propagation speeds and directions
making difficult to separate magnetic and temperature effects (see
\eg\ \cite{Moradi+Cally2008}). To understand the influence of the
magnetic field on travel time measurements, forward modeling of
waves in magnetic regions is required. This has become the
preferred approach in recent years.

With this aim, we performed 2D numerical simulations of
magneto-acoustic wave propagation though a series of model
sunspots with different field strength, from the deep interior to
chromospheric layers \cite{Khomenko+Collados2008}. The waves are
excited by an external force localized in space just below the
photosphere at $-200$ km, according to the models of wave
excitation in the Sun \cite{Nordlund+Stein2001,
Stein+Nordlund2001}. The spectral properties of the source
resemble the spectrum of solar waves with the maximum power at 3.3
mHz. In the experiment described below the source is located at 12
Mm far from sunspot axis in the region where the acoustic speed is
slightly larger than the Alfv\'en speed. The details of the
numerical procedure are given in \cite{Khomenko+etal2008b}.

The magneto-static sunspot models are calculated using the method
proposed in \cite{Khomenko+Collados2008}. The sunspot models have
a cool zone below the surface down to, about, $-2$ Mm depth. Below
this depth the temperature gradient in horizontal direction is
small and no hot zone is introduced in the present study. The
photospheric field strength in three models used is of $B_{\rm
phot}$=0.9, 1.5 and 2.4 kG.

Fig.~\ref{fig:magnetic} shows a snapshot of the simulations. The
fast magneto-acoustic modes (analog of $p$-modes in the quiet Sun)
can be distinguished propagating below the surface in the pressure
and velocity variations. In addition, there is a perturbation with
much smaller vertical wavelength visible best in the magnetic
field and transversal velocity variations. A part of this
perturbation is a slow MHD mode generated directly by the source
at horizontal position $X=-12$ Mm. This mode propagates with a
visibly low speed downwards along the sunspot magnetic field
lines. In addition to the slow MHD waves generated directly by the
source there is another wave type. The  propagation speed of this
small vertical wavelength disturbance is comparable to that of the
fast modes. Unlike the slow MHD waves, these waves propagate
horizontally across the sunspot. We conclude that these waves are
magneto-gravity waves. The variations produced by these
magneto-gravity waves decrease rapidly with depth and disappear
almost completely below $-3$ Mm (top left panel of
Fig.~\ref{fig:magnetic}).

%Sunspot structure - cold zone below the photosphere down to 3 Mm.
%No hot zone. Acoustic waves should lag behind their quiet Sun
%counterparts. But they instead speed up due to magnetic field. The
%effect depends on the field strength. New modes propagate below
%the surface and are difficult to detect in the photosphere.
%Spectral and spatial properties of a source.

%%%%%%%%%%%%%%%%%%%%%%%%%%%%%%%%%%%%%%%%%%%%%%%%%%%%%%%%%%%%%%%%%%%%%%%%%%%%%%%%%%%%%%
\begin{figure}[t]
\includegraphics[width=6.7cm]{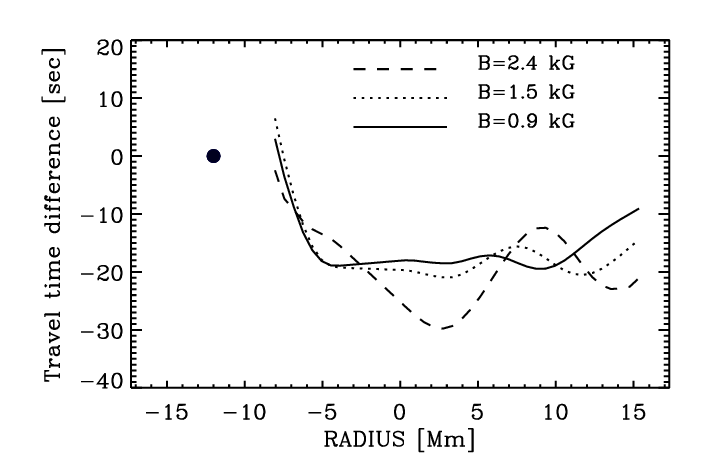}
\sidecaption \caption{Phase travel-time difference calculated
between the travel times measured in simulations without magnetic
field and simulations with different model sunspots (indicated in
the figure) as a function of horizontal distance. The source
location at $X_0=$ 12 Mm to the left of the sunspot axis is marked
by a circle.} \label{fig:traveltime}
\end{figure}
%%%%%%%%%%%%%%%%%%%%%%%%%%%%%%%%%%%%%%%%%%%%%%%%%%%%%%%%%%%%%%%%%%%%%%%%%%%%%%%%%%%%%%%

What are the surface signatures of these modes and how do they
affect the travel time measurements in solar active regions? As
follows from the bottom right panel of Fig.~\ref{fig:magnetic}, in
the photospheric layers the dominant variations of the
longitudinal (vertical) velocity are due to the fast
magneto-acoustic mode propagating horizontally across the sunspot.
Fig.~\ref{fig:traveltime} gives the travel time differences
between the phase travel times of the this mode measured in the
sunspot photosphere relative to the non-magnetic quiet Sun. The
latter is represented by standard solar model S
\cite{Christensen-Dalsgaard+etal1996}.  The travel times are
obtained from a Gabor's wavelet fit to the simulated time-distance
diagrams. Negative values mean that waves in the magnetic
simulations propagate faster. Here, it must be recalled that the
model sunspots have a cool zone below the photosphere implying a
lower acoustic speed. Thus, if the waves were purely acoustic in
nature the travel time differences in Fig.~\ref{fig:traveltime}
would have positive sign. Instead the propagation speed of the
fast magneto-acoustic waves increases with the magnetic field,
which has the natural consequences observed in
Fig.~\ref{fig:traveltime}: the waves in sunspot models with larger
magnetic field propagate faster. The values of the travel times
differences that we obtain from simulations agree rather well with
those measured in solar active regions (see \eg\
\cite{Couvidat+Birch+Kosovichev2006}). Thus, we conclude that the
wave propagation below solar active regions is governed by the
magnetic field. Despite the new wave modes generated in the
sunspot atmosphere do not affect directly the time-distance
helioseismology measurements, the travel times of the fast MHD
modes (analog of the $p$-modes) are affected by the magnetic
field. A more complete analysis of these simulations is presented
in \cite{Khomenko+etal2008b}.

%%%%%%%%%%%%%%%%%%%%%%%%%%%%%%%%%%%%%%%%%%%%%%%%%%%%%%%%%%%%%%%%%%%%%%%%%%%%%%%%%%%%%%
\begin{figure}[b]
\includegraphics[width=12cm]{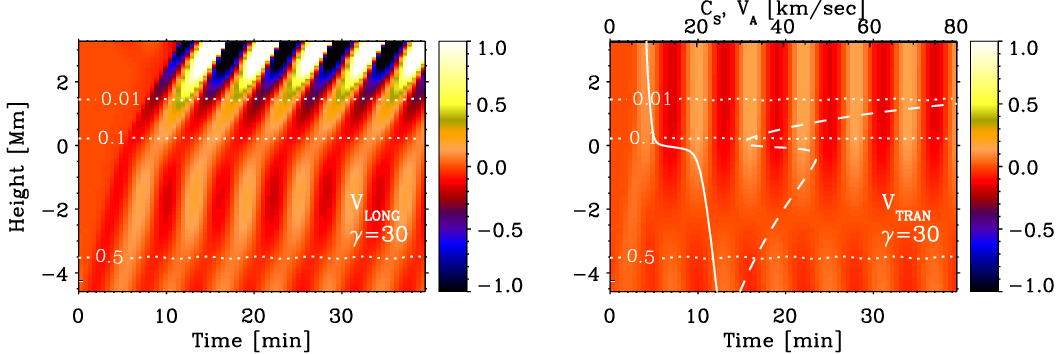}
\caption{Variations of the longitudinal (left) and transversal
(right) velocities with height and time in a roAp star with
$T_{\rm eff}=$7750 K and log$g$=4.0. The units are km/sec. Results
are for a dipolar magnetic field strength $B=1000$ G, at a
latitude where the inclination with respect to the local vertical
is 30 degrees and oscillation frequency 2.8 mHz. Zero height
corresponds to the bottom photosphere. Contours of constant
$c_S^2/v_A^2$ are plotted with dotted lines. Height dependences of
the $c_S$ (solid line) and $v_A$ (dashed line) are plotted over
the right panel, the scale is given by the upper axis.}
\label{fig:timeheight}
\end{figure}
%%%%%%%%%%%%%%%%%%%%%%%%%%%%%%%%%%%%%%%%%%%%%%%%%%%%%%%%%%%%%%%%%%%%%%%%%%%%%%%%%%%%%%%

\section{Oscillations in magnetic roAp stars}

Closely related to the problems of local helioseismology in
magnetic regions is the problem of interpreting the oscillations
in magnetic rapidly oscillating peculiar Ap stars. These stars
posses a strong dipolar-like magnetic field of 1--25 kG and
horizontal and vertical stratification of chemical composition.
They pulsate with periods between 4 and 20 minutes. This offers a
unique opportunity to study the interaction between convection,
waves and strong magnetic field. Recent reviews on the properties
of these stars and their pulsations can be found in Kurtz
\cite{Kurtz2008}, Cunha \cite{Cunha2007} and Kochukhov
\cite{Kochukhov2007, Kochukhov2008}.

Several properties of the oscillations observed on these stars
make them peculiar. The amplitudes of the pulsations vary with
rotation period according to the magnetic field structure giving
rise to oblique pulsator model \cite{Kurtz1982}. The spectral
lines of different chemical elements pulsate with order of
magnitude different amplitudes and significant phase shifts
between them, depending on their formation heights
\cite{Ryabchikova+etal2002}. In addition, several stars pulsate
with frequencies exceeding the acoustic cut-off predicted by
stellar models \cite{Sousa+Cunha2008}.
Due to their strong magnetic fields, the atmospheres of roAp stars
are regions where the magnetic pressure exceeds the gas pressure
and the oscillations are magnetically dominated. Recent analytical
modeling of high-frequency waves \cite{Sousa+Cunha2008} suggest
that the pulsations observed in the atmospheric layers can be a
superposition of running acoustic waves (slow MHD) and nearly
standing magnetic waves (fast MHD) that are nearly decoupled in
the region $\beta \ll 1$.

In order to identify the wave modes observed in the atmospheres of
roAp stars, we solved numerically the governing non-linear MHD
equations in 2D geometry for a semi-empirical model atmosphere.
The model has $T_{\rm eff}=$7750 K and log$g$=4.0.
We assumed that: (1) magnetic field varies on spatial scales much
larger than the typical wavelength, allowing the problem to be
solved locally for a plane-parallel atmosphere with a homogeneous
inclined magnetic field; (2) waves in the atmosphere are excited
by low-degree pulsation modes with radial velocities exceeding
horizontal velocities. We studied a grid with the magnetic field
strength $B$ varying from 1 to 7 kG, its inclination varying
between 0 and 60 degrees and pulsation frequencies between 1.25
and 2.8 mHz (below and above the cut-off frequency of this
simulated star).

%%%%%%%%%%%%%%%%%%%%%%%%%%%%%%%%%%%%%%%%%%%%%%%%%%%%%%%%%%%%%%%%%%%%%%%%%%%%%%%%%%%%%%
\begin{figure}
\center
\includegraphics[width=10cm]{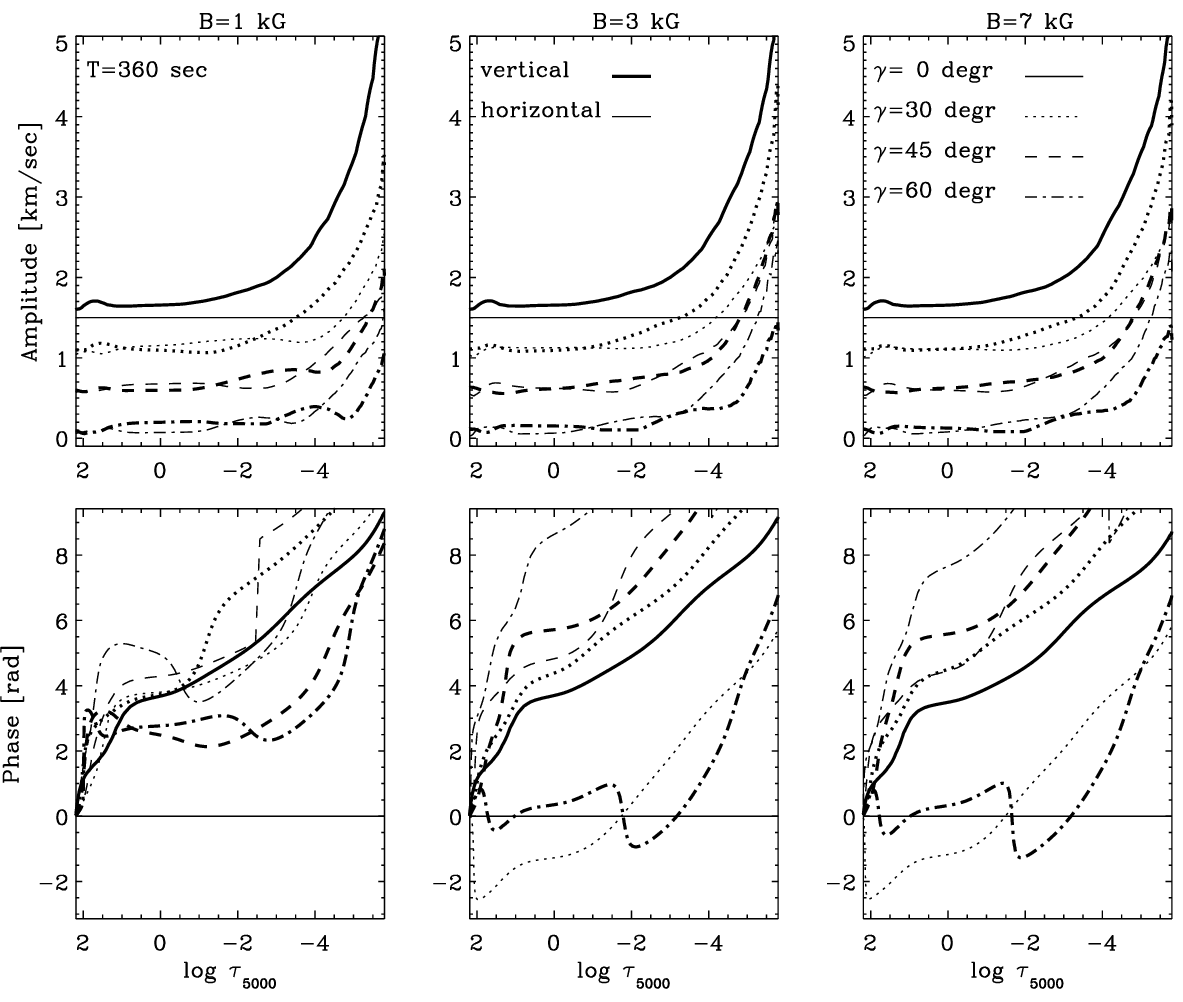}
\caption{Amplitudes (top) and phases (bottom) of the vertical
(thick lines) and horizontal (thin lines) velocities as a function
of optical depth in the atmosphere of a roAp star obtained from
the simulations with dipolar magnetic field of different strength
$B$ and at different latitudes corresponding to inclinations
$\gamma$ (indicated in the figure). The results are for the
oscillation frequency 2.8 mHz. The amplitude curves for each
$\gamma$ are separated by 0.5 km/sec for better visualization.}
\label{fig:roap}
\end{figure}
%%%%%%%%%%%%%%%%%%%%%%%%%%%%%%%%%%%%%%%%%%%%%%%%%%%%%%%%%%%%%%%%%%%%%%%%%%%%%%%%%%%%%%%

Despite the simple magnetic field geometry, the simulations give
rise to a complex picture of the superposition of several modes,
varying significantly depending on the parameters of the
simulations. An example of the velocity field developed in the
simulation with $B=1$ kG, $\gamma=30$ degrees and pulsation period
$T_0=360$ sec is given in Fig.~\ref{fig:timeheight}. Longitudinal
and transversal projections of the velocity with respect to the
local magnetic field allow us to separate clearly the wave modes.
The longitudinal velocity component shows the presence of the slow
MHD (acoustic) wave. Under $\beta \ll 1$ conditions, this wave is
propagating along the inclined magnetic field. The transversal
velocity reveals the fast MHD (magnetic) wave. The rapid increase
of the Alfv\'en speed with height makes the wavelength of this
mode extremely large occupying the whole atmosphere. While at the
bottom layers (below the photosphere) the amplitudes of the fast
and slow modes are comparable, in the upper atmosphere the slow
mode clearly dominates since its amplitude increases exponentially
with height, much more that that of the fast mode. Two node
heights can be observed in the case of the slow mode (at $-3.5$
and $0$ Mm) and one node height in the case of the fast mode (at
$-2$ Mm), all produced by wave reflection. The model atmosphere
used in the simulations has strong density and temperature jumps
at the photospheric level, producing efficient reflection. We can
observe an evanescent pattern of the slow mode at heights between
$-3$ and $0$ Mm (left panel of Fig.~\ref{fig:timeheight}). Below
and above these heights the slow wave is propagating with a speed
defined by the local speed of sound. One can appreciate a
considerably slower propagation in the upper layers due to the
smaller sound speed. Strong slow wave shocks are formed above 2 Mm
height with amplitudes up to 5 km/sec. The amplitudes of the
velocities obtained in the simulations are similar to those
observed \cite{Kochukhov2007}.

Fig.~\ref{fig:roap} gives the amplitudes and phases of the
horizontal and vertical velocities as a function of the optical
depth for different field strengths and inclinations obtained in
the simulations with the pulsation period of 360 sec. The
superposition of the fast and slow waves produces additional
node-like structures at heights where these modes interfere
destructively. Fig.~\ref{fig:roap} shows that both amplitudes and
phases of the velocity are complex functions of optical depth and
of the parameters of the simulations. In general, the amplitude of
the vertical velocity decreases with the inclination, while the
amplitude of the horizontal velocity increases. In the case of the
inclination $\gamma \ne 0$, the amplitudes of the waves at the top
of the atmosphere are smaller for $B=1$ kG compared to larger
field strengths. This can be explained by the decrease of the
cut-off frequency due to preferred wave propagation in the
direction of the magnetic field \cite{Roberts1983}. The location
of the node-like surfaces and waves propagating up and down at
different heights can be appreciated from the phase curves at the
bottom panels of Fig.~\ref{fig:roap}. All these features are
similar to observations.

The disc-integrated velocity signal produced by the atmospheric
pulsations of such a star would depend in a complex way on the
inclination of the magnetic axis with respect to the observational
line of sight and needs a further study. However, we can conclude
that the velocity signal observed in the upper atmospheric layers
of roAp stars is mostly due to running slow mode acoustic waves.
The node structures and the rapid phase variations at the lower
atmospheric layers are due to multiple reflections and
interference of the slow and fast MHD wave modes.

%One figure, atmospheric influence on waves, simulation grid.

\section{Conclusions}

Magnetic field introduces an additional restoring force and makes
the propagation of waves in stellar atmospheres more complex
compared to the case of acoustic-gravity waves. We have developed
a numerical code that allows the modeling of waves inside magnetic
fields for a large variety of phenomena, from Sun to stars.
We have applied this code to study the role of different solar
magnetic structures concerning wave energy transport to the upper
atmosphere; interpretation of local helioseismology measurements
in solar active regions; pulsations of magnetic roAp stars.
Puzzling physics of the interaction of waves with the magnetic
field in a variety of magnetic field configurations in the Sun and
stars is to be explored in the future.

%\fullreferences


\begin{thebibliography}{}

\bibitem{Bi+Li1998}
Bi, S. and Li, R.: 1998,
\newblock ``Excitation of the solar $p$-modes by turbulent stress'', {\em
  Astron.\ Astrophys.\/} {\bf 335}, 673---678

\bibitem{Birch+Kosovichev2000}
Birch, A.~C. and Kosovichev, A.~G.: 2000,
\newblock ``Travel Time Sensitivity Kernels'', {\em Solar Phys.\/} {\bf 192},
  193---201

\bibitem{Bloomfield+etal2007}
    Bloomfield, D. S., Lagg, A., and Solanki, S. K.: 2007,
\newblock ``The Nature of Running Penumbral Waves Revealed'', {\em Astrophys.\ J.\/} {\bf 671},
  1005---1012

\bibitem{Bogdan+etal2003}
Bogdan, T.~J., Carlsson, M., Hansteen, V., McMurry, A., Rosenthal, C.~S.,
  Johnson, M., Petty-Powell, S., Zita, E.~J., Stein, R.~F., McIntosh, S.~W.,
  and Nordlund, A.: 2003,
\newblock ``Waves in the magnetized solar atmosphere. II. Waves from localized
  sources in magnetic flux concentrations'', {\em Astrophys.\ J.\/} {\bf 599},
  626---660

\bibitem{Centeno+etal2006b}
Centeno, R., Collados, M., and \mbox{Trujillo Bueno}, J.: 2006,
\newblock ``Oscillations and Wave Propagation in Different Solar Magnetic
  Features'', in R. Casini and B.~W. Lites (Eds.), {\em Solar Polarization
  4\/}, Vol. 358, ASP Conference Series,  465

\bibitem{Christensen-Dalsgaard+etal1996}
Christensen-Dalsgaard, J., Dappen, W., Ajukov, S.~V., and \mbox{30 co-authors}:
  1996,
\newblock ``The Current State of Solar Modeling'', {\em Science\/} {\bf 272},
  1286

\bibitem{Couvidat+Birch+Kosovichev2006}
Couvidat, S., Birch, A.~C., and Kosovichev, A.~G.: 2006,
\newblock ``Three-dimensional Inversion of Sound Speed below a Sunspot in the
  Born Approximation'', {\em Astrophys.\ J.\/} {\bf 640}, 516---524

\bibitem{Cunha2007}
Cunha, M.~S.: 2007,
\newblock ``Theory of rapidly oscillating Ap stars'', {\em Communications in
  Astroseismology\/} {\bf 150}, 48---54

\bibitem{Deubner1975}
Deubner, F.-L.: 1975,
\newblock ``Some properties of velocity fields in the solar photosphere. V -
  Spatio-temporal analysis of high resolution spectra'', {\em Solar Phys.\/}
  {\bf 39}, 31---48

\bibitem{Duvall+etal1993}
Duvall, T. L.~J., Jefferies, S.~M., Harvey, J.~W., and Pomerantz, M.~A.: 1993,
\newblock ``Time-distance helioseismology'', {\em Nature\/} {\bf 362},
  430---432

\bibitem{Fleck+Schmitz1991}
Fleck, B. and Schmitz, F.: 1991,
\newblock ``The 3-min oscillations of the solar chromosphere - A basic physical
  effect?'', {\em Astron.\ Astrophys.\/} {\bf 250}, 235---244

\bibitem{Gizon+Birch2002}
Gizon, L. and Birch, A.~C.: 2002,
\newblock ``Time-Distance Helioseismology: The Forward Problem for Random
  Distributed Sources'', {\em Astrophys.\ J.\/} {\bf 571}, 966---986

\bibitem{Goldreich+Kumar1990}
Goldreich, P. and Kumar, P.: 1990,
\newblock ``Wave generation by turbulent convection'', {\em Astrophys.\ J.\/}
  {\bf 363}, 694---704

\bibitem{Goldreich+Murray+Kumar1994}
Goldreich, P., Murray, N., and Kumar, P.: 1994,
\newblock ``Excitation of solar p-modes'', {\em Astrophys.\ J.\/} {\bf 424},
  466---479

\bibitem{Hasan+Ballegooijen2008}
Hasan, S.~S. and \mbox{van Ballegooijen}, A.~A.: 2008,
\newblock ``Dynamics of the Solar Magnetic Network. II. Heating the Magnetized
  Chromosphere'', {\em Astron.\ Astrophys.\/} {\bf 680}, 1542---1552

\bibitem{Hasan+etal2005}
Hasan, S.~S., \mbox{van Ballegooijen}, A.~A., Kalkofen, W., and Steiner, O.:
  2005,
\newblock ``Dynamisc of solar magnetic network: two-dimensional MHD
  simulations'', {\em Astrophys.\ J.\/} {\bf 631}, 1270---1280

\bibitem{Hindman+Jain2008}
Hindman, B.~W. and Jain, R.: 2008,
\newblock ``The Generation of Coronal Loop Waves below the Photosphere by
  p-Mode Forcing'', {\em Astrophys.\ J.\/} {\bf 677}, 769---780

\bibitem{Jacoutot+etal2008}
Jacoutot, L., Kosovichev, A.~G., Wray, A., and Mansour, N.~N.: 2008,
\newblock ``Realistic Numerical Simulations of Solar Convection and
  Oscillations in Magnetic Regions'', {\em Astrophys.\ J.\/} {\bf 684},
  L51---L54

\bibitem{Kalkofen2007}
Kalkofen, W.: 2007,
\newblock ``Is the Solar Chromosphere Heated by Acoustic Waves?'', {\em
  Astrophys.\ J.\/} {\bf 671}, 2154---2158

\bibitem{Khomenko+etal2008}
Khomenko, E., Centeno, R., Collados, M., and \mbox{Trujillo Bueno}, J.: 2008a,
\newblock ``Channeling 5 Minute Photospheric Oscillations into the Solar Outer
  Atmosphere through Small-Scale Vertical Magnetic Flux Tubes'', {\em
  Astrophys.\ J.\/} {\bf 676}, L85---L88

\bibitem{Khomenko+Collados2006}
Khomenko, E. and Collados, M.: 2006,
\newblock ``Numerical Modeling of Magnetohydrodynamic Wave Propagation and
  Refraction in Sunspots'', {\em Astrophys.\ J.\/} {\bf 653}, 739---755

\bibitem{Khomenko+Collados2008}
Khomenko, E. and Collados, M.: 2008,
\newblock ``MHS sunspot model from deep sub-photospheric to chromospheric
  layers'', {\em Astrophys.\ J.\/} in press, arXiv:0808.3571

\bibitem{Khomenko+Collados2007}
Khomenko, E., Collados, M., and Feliipe, T.: 2008b,
\newblock ``Non-linear numerical simulations of magneto-acoustic wave
  propagation in small-scale flux tubes'', {\em Solar Phys.\/} {\bf 251},
  589---611

\bibitem{Khomenko+etal2008b}
Khomenko, E., Kosovichev, A., Collados, M., Parchevsky, K., and Olshevsky, V.:
  2008c,
\newblock ``Theoretical modeling of propagation of magneto-acoustic waves in
  magnetic regions below sunspots'', {\em Astrophys.\ J.\/}  submitted,
  arXiv:0809.0278

\bibitem{Kochukhov2007}
Kochukhov, O.: 2007,
\newblock ``Observations of pulsations in roAp stars'', {\em Communications in
  Astroseismology\/} {\bf 150}, 39---47

\bibitem{Kochukhov2008}
Kochukhov, O.: 2008,
\newblock ``Pulsation in the atmosphere of roAp stars'', {\em Communications in
  Astroseismology\/} arXiv:0810.1508, in press

\bibitem{Kosovichev1999}
Kosovichev, A.~G.: 1999,
\newblock ``Inversion methods in helioseismology and solar tomography'', {\em
  J. Comput. Appl. Math\/} {\bf 109}, 1---39

\bibitem{Kosovichev2002}
Kosovichev, A.~G.: 2002,
\newblock ``Subsurface structure of sunspots'', {\em Astron.\ Nachr.\/} {\bf
  323}, 186---191

\bibitem{Kosovichev+Duvall1997}
Kosovichev, A.~G. and Duvall, T. L.~J.: 1997,
\newblock ``Acoustic tomography of solar convective flows and structures'', in
  F. Pijpers, J. Christensen-Dalsgaard, and C. Rosenthal (Eds.), {\em Solar
  Convection and Oscillations and their Relationship\/}, Vol. 225, Astrophysics
  and Space Science Library, Kluwer Academic Publishers,  241---260

\bibitem{Kosovichev+etal2000}
Kosovichev, A.~G., Duvall, T. L.~J., and Scherrer, P.~H.: 2000,
\newblock ``Time-Distance Inversion Methods and Results'', {\em Solar Phys.\/}
  {\bf 192}, 159---176

\bibitem{Krijer+etal2001}
Krijger, J.~M., Rutten, R.~J., Lites, B.~W., Straus, T., Shine, R.~A., and
  Tarbell, T.~D.: 2001,
\newblock ``Dynamics of the solar chromosphere. III. Ultraviolet brightness
  oscillations from TRACE'', {\em Astron.\ Astrophys.\/} {\bf 379}, 1052---1082

\bibitem{Kurtz1982}
Kurtz, D.~W.: 1982,
\newblock ``Rapidly oscillating AP stars'', {\em Mon.\ Not.\ R.\ Astron.\
  Soc.\/} {\bf 200}, 807---859

\bibitem{Kurtz2008}
Kurtz, D.~W.: 2008,
\newblock ``The Solar-Stellar Connection: Magneto-Acoustic Pulsations in 1.5 M
  o. – 2 M o. Peculiar A Stars'', {\em Solar Phys.\/} {\bf 251}, 21---30

\bibitem{Lighthill1952}
Lighthill, M.~J.: 1952,
\newblock ``On sound generated aerodynamically. I. General theory'', {\em Proc.
  R. Soc. London A.\/} {\bf 211}, 564---587

\bibitem{DeMoortel+etal2002}
\mbox{De Moortel}, I., Ireland, J., Hood, A.~W., and Walsh, R.~W.: 2002,
\newblock ``The detection of 3 and 5 min period oscillations in coronal
  loops'', {\em Astron.\ Astrophys.\/} {\bf 387}, L13---L16


\bibitem{Moradi+Cally2008}
Moradi, H. and Cally, P.: 2008,
\newblock ``Time-Distance modelling in a simulated sunspot atmosphere'', {\em
  Solar Phys.\/} in press

\bibitem{Moradi+etal2008}
Moradi, H., Hanasoge, S., and Cally, P.~S.: 2008,
\newblock ``Numerical Models of Travel Time Inhomogeneities in Sunspots'', {\em
  Astrophys.\ J.\/} submitted, arXiv:0808.3628

\bibitem{Musielak+etal1994}
Musielak, Z.~E., Rosner, R., Stein, R.~F., and Ulmschneider, P.: 1994,
\newblock ``On sound generation by turbulent convection: A new look at old
  results'', {\em Astrophys.\ J.\/} {\bf 423}, 474---487

\bibitem{Nagashima+etal2007}
Nagashima, K., Sekii, T., Kosovichev, A.~G., Shibahashi, H., Tsuneta, S.,
  Ichimoto, K., Katsukawa, Y., Lites, B., Nagata, S., Shimizu, T., Shine,
  R.~A., Suematsu, Y., Tarbell, T.~D., and Title, A.~M.: 2007,
\newblock ``Observations of Sunspot Oscillations in G Band and CaII H Line with
  Solar Optical Telescope on Hinode'', {\em Publ.\ Astron.\ Soc.\ Pac.\/} {\bf
  59}, S631---S636

\bibitem{Nordlund+Stein2001}
Nordlund, {\AA}. and Stein, R.~F.: 2001,
\newblock ``Solar Oscillations and Convection. I. Formalism for Radial
  Oscillations'', {\em Astrophys.\ J.\/} {\bf 546}, 576---584

\bibitem{Osterbrock1961}
Osterbrock, D.~E.: 1961,
\newblock ``The heating of the solat chromosphere, plages and corona by
  magnetohydrodynamis waves'', {\em Astrophys.\ J.\/} {\bf 134}, 347---388

\bibitem{Pizzo1986}
Pizzo, V.~J.: 1986,
\newblock ``Numerical solution of the magntostatic equations for thick flux
  tubes, with application to sunspots, pores and related structures'', {\em
  Astrophys.\ J.\/} {\bf 302}, 785---808

\bibitem{Pneuman+etal1986}
Pneuman, G.~W., Solanki, S.~K., and Stenflo, J.~O.: 1986,
\newblock ``Structure and merging of solar magnetic fluxtubes'', {\em Astron.\
  Astrophys.\/} {\bf 154}, 231---242

\bibitem{DePontieu+etal2003}
\mbox{De Pontieu}, B., Erdelyi, R., and de~Wijn, A.~G.: 2003,
\newblock ``Intensity Oscillations in the Upper Transition Region above Active
  Region Plage'', {\em Astrophys.\ J.\/} {\bf 595}, L63---L66

\bibitem{DePontieu+etal2004}
\mbox{De Pontieu}, B., Erdelyi, R.~J., and Stewart, P.: 2004,
\newblock ``Solar chromospheric spicules from the leakage of photospheric
  oscillations and flows'', {\em Nature\/} {\bf 430}, 536---539


\bibitem{Roberts1983}
Roberts, B.: 1983,
\newblock ``Wave propagation in intense flux tubes'', {\em Solar Phys.\/} {\bf
  87}, 77---93

\bibitem{Rosenthal+etal2002}
Rosenthal, C.~S., Bogdan, T.~J., Carlsson, M., Dorch, S. B.~F., Hansteen, V.,
  McIntosh, S.~W., McMurry, A., Nordlund, A., and Stein, R.~F.: 2002,
\newblock ``Waves in the magnetized solar atmosphere. I. Basic processes and
  internetwork oscillations'', {\em Astrophys.\ J.\/} {\bf 564}, 508---524

\bibitem{Rouppe+etal2003}
Rouppe van~der Voort, L. H. M.~R., Rutten, R.~J., S{\"u}tterlin, P., Sloover, P.~J.,
  and Krijger, J.~M.: 2003,
\newblock ``La Palma observations of umbral flashes'', {\em Astron.\
  Astrophys.\/} {\bf 403}, 277---285


\bibitem{Ryabchikova+etal2002}
Ryabchikova, T., Piskunov, N., Kochukhov, O., Tsymbal, V., Mittermayer, P., and
  Weiss, W.~W.: 2002,
\newblock ``Abundance stratification and pulsation in the atmosphere of the
  roAp star boldmath gamma Equulei'', {\em Astron.\ Astrophys.\/} {\bf 384},
  545---553

\bibitem{Solanki2003}
Solanki, S.~K.: 2003,
\newblock ``The sunspots'', {\em Astron.\ Astrophys.\ Rev.\/} {\bf 11},
  153--286

\bibitem{Sousa+Cunha2008}
Sousa, S.~G. and Cunha, M.~S.: 2008,
\newblock ``On mode conversion and wave reflection in magnetic Ap stars'', {\em
  Mon.\ Not.\ R.\ Astron.\ Soc.\/} {\bf 386}, 531---542

\bibitem{Stein1967}
Stein, R.: 1967,
\newblock ``Generation of Acoustic and Gravity Waves by Turbulence in an
  Isothermal Stratified Atmosphere'', {\em Solar Phys.\/} {\bf 2}, 385---432

\bibitem{Stein+Nordlund2001}
Stein, R.~F. and Nordlund, {\AA}.: 2001,
\newblock ``Solar oscillations and convection. II. Excitation of radial
  oscillations'', {\em Astrophys.\ J.\/} {\bf 546}, 585---603

\bibitem{Tziotziou+etal2006}
Tziotziou, K., Tsiropoula, G., Mein, N., and Mein, P.: 2006,
\newblock ``Observational characteristics and association of umbral
  oscillations and running penumbral waves'', {\em Astron.\ Astrophys.\/} {\bf
  456}, 689---695

\bibitem{Tziotziou+etal2007}
Tziotziou, K., Tsiropoula, G., Mein, N., and Mein, P.: 2007,
\newblock ``Dual-line spectral and phase analysis of sunspot oscillations'',
  {\em Astron.\ Astrophys.\/} {\bf 463}, 1153---1163

\bibitem{Ulmschneider1971}
Ulmschneider, P.: 1971,
\newblock ``On the propagation of a spectrum of acoustic waves in the solar
  atmosphere'', {\em Astron.\ Astrophys.\/} {\bf 14}, 275---282

\bibitem{Ulrich1970}
Ulrich, R.~K.: 1970,
\newblock ``The Five-Minute Oscillations on the Solar Surface'', {\em
  Astrophys.\ J.\/} {\bf 162}, 993---1002


\bibitem{Veccio+etal2007}
Vecchio, A., Cauzzi, G., Reardon, K.~P., Janssen, K., and Rimmele, T.: 2007,
\newblock ``Solar atmospheric oscillations and the chromospheric magnetic
  topology'', {\em Astron.\ Astrophys.\/} {\bf 461}, L1---L4

\bibitem{Wunnenberg+etal2002}
Wunnenberg, M., Kneer, F., and Hirzberger, J.: 2002,
\newblock ``Evidence for short-period acoustic waves in the solar atmosphere'',
  {\em Astron.\ Astrophys.\/} {\bf 395}, L51---L54

\bibitem{Zhao+Kosovichev2003}
Zhao, J. and Kosovichev, A.~G.: 2003,
\newblock ``Helioseismic Observation of the Structure and Dynamics of a
  Rotating Sunspot Beneath the Solar Surface'', {\em Astrophys.\ J.\/} {\bf
  591}, 446---453

\end{thebibliography}
\end{document}